# Deterministically fabricated strain-tunable quantum dot single-photon sources emitting in the telecom O-band

N. Srocka[1], P. Mrowiński[1,2], J. Große[1], M. Schmidt[1,3], S. Rodt[1], S. Reitzenstein[1]

[1]*Institut für Festkörperphysik, Technische Universität Berlin, Hardenbergstraße 36, D-10623 Berlin, Germany*

[2]*Laboratory for Optical Spectroscopy of Nanostructures, Department of Experimental Physics, Wrocław University of Technology, Wybrzeże Wyspiańskiego 27, Wrocław, Poland*

[3]*Physikalisch-Technische Bundesanstalt, Abbestraße 2-12, 10587 Berlin, Germany*

**Abstract**

Most quantum communication schemes aim at the long-distance transmission of quantum information. In the quantum repeater concept, the transmission line is subdivided into shorter links interconnected by entanglement distribution via Bell-state measurements to overcome inherent channel losses. This concept requires on-demand single-photon sources with a high degree of multi-photon suppression and high indistinguishability within each repeater node. For a successful operation of the repeater, a spectral matching of remote quantum light sources is essential. We present a spectrally tunable single-photon source emitting in the telecom O-band with the potential to function as a building block of a quantum communication network based on optical fibers. A thin membrane of GaAs embedding InGaAs quantum dots (QDs) is attached onto a piezoelectric actuator via gold thermocompression bonding. Here the thin gold layer acts simultaneously as an electrical contact, strain transmission medium and broadband backside mirror for the QD-micromesa. The nanofabrication of the QD-micromesa is based on in-situ electron-beam lithography, which makes it possible to integrate pre-selected single QDs deterministically into the center of monolithic micromesa structures. The QD pre-selection is based on distinct single-QD properties, signal intensity and emission energy. In combination with strain-induced fine tuning this offers a robust method to achieve spectral resonance in the emission of remote QDs. We show that the spectral tuning has no detectable influence on the multi-photon suppression with $g^{(2)}(0)$ as low as 2-4% and that the emission can be stabilized to an accuracy of 4 μeV using a closed-loop optical feedback.

The emission of single photons with controllable wavelength and high indistinguishability is a key parameter in quantum light sources for quantum nanophotonics. It is, for example, required to implement boson sampling experiments[1] and to realize entanglement swapping via Bell-state measurements in large-distance quantum communication networks based on the quantum repeater concept.[2–4] In fact, in case of the quantum repeater it is crucial to fabricate a chain of wavelength-tunable quantum light sources that provide identical photons on demand. Even more, it is important to operate such light sources at telecommunication wavelengths, i.e. within the telecom O-band (~1.3 μm) or C-band (~1.55 μm) with enhanced brightness[5] and deterministic fabrication scheme,[6] to pave the way towards the real-world implementation of long-distance quantum communication networks via optical fibers, as recently reported using electrical Stark tuning of a QD device.[7]

In this work, we demonstrate technological advances and experimental findings to realize wavelength-tunable quantum emitters of high single-photon purity in the telecom O-band. The strain-tunable emitters are based on self-assembled InGaAs quantum dots (QDs) that are deterministically integrated into photonic nanostructures attached to piezo-elements by means of a flip-chip process.[8,9] Here, the emission range of the InGaAs QDs is redshifted to the telecom O-band by using a strain-reducing layer (SRL),[10,11] while the piezoelectric actuator allows us to apply an external strain field to fine-tune and stabilize the QD emission wavelength during operation of the quantum light sources. To demonstrate their application potential, we examined the optical properties of two O-band single-QD mesa structures. The emission of their charged excitonic (CX) states, which are relevant for efficient single-photon generation,[12] are first studied regarding multi-photon suppression. Secondly, we test the wavelength tuning capabilities via strain-tuning and we introduce active spectral stabilization based on a closed-loop proportional-integral-derivative (PID) controller. Finally, we demonstrate strain-tuning and active stabilization of two remote QDs, which is a crucial step over existing results towards the real-world application of QD-SPS in advanced quantum communication schemes where it can pave the way towards two-photon interference (TPI) of remote sources in the telecom O-band required for the implementation of fiber-based quantum repeater networks.[4]

We manufactured our device in three main technological steps: i) growing a semiconductor heterostructure by metal organic chemical vapor deposition, ii) gold bonding a few-hundreds nm-thin membrane of this structure onto a piezoelectric actuator, and iii) nanostructuring the QD-membrane deterministically by low-temperature cathodoluminescence (CL) scanning and in-situ EBL.

First, a 200 nm GaAs buffer is grown on n-doped GaAs (100) substrate followed by 1 µm thick $Al_{0.90}Ga_{0.10}As$ etch stop layer, a 2 µm thick GaAs layer and another 100 nm thick $Al_{0.90}Ga_{0.10}As$ layer. These three layers and the substrate are sacrificial layers to be removed during post-growth processing. The final 879 nm thick region of GaAs forms the active device membrane and includes a single InGaAs QD layer combined with the SRL, with the QD layer located 637 nm above the second etch stop layer (see Fig. 1a). The QD layer is formed by 1.5 monolayers of $In_{0.7}Ga_{0.3}As$ followed by a 0.5 monolayer GaAs flush. The subsequent 5.5 nm thick InGaAs SRL has a gradual decrease of the In-content from 30% to 10% over the first 3.5 nm.

Next, the as-grown sample and a PIN-PMN-PT ($Pb(In_{1/2}Nb_{1/2})O_3$-$Pb(Mg_{1/3}Nb_{2/3})O_3$-$PbTiO_3$) piezoelectric crystal are sputtered with 250 nm of gold. Their surfaces are then activated under argon plasma and, immediately after being removed from the deposition tool, bonded face-to-face under ambient air and thermocompression. In this step, a pressure of 6 MPa and a temperature of about 600 K are applied. This process allows for a stable bond that is not affected by the inherent surface roughness of the PIN-PMN-PT crystal. For more details on this process, see Ref. [13].

The flip-chip process is finalized by a layer-by-layer wet-etching process. First, the 400 µm thick GaAs substrate is removed by a fast and aggressive etchant ($H_2O_2/NH_3$, 10:1), stopped at the first $Al_{0.9}Ga_{0.1}As$ layer due to the high selectivity of the wet-chemical etching process.[14] HCL acid is then used to solely remove the exposed etch stop layer. In our process we use HCl instead of the typically applied HF mainly because of the lower etch rate which leads to better process control and higher quality of the etched surfaces. The now exposed second and thinner GaAs layer (2 µm) is lifted off by a slower etchant (citric acid/$H_2O_2$, 4:1), which leads to a significantly improved surface roughness if compared to a single-step etch process. Again, the second etch stop layer is removed by HCL acid. At this point an 885 nm thick membrane, containing the single QD layer, is gold bonded to a PIN-PMN-PT crystal and completely freed from sacrificial layers.

In preparation for the following main processing step the sample is spin coated with a nominal 300 nm thick layer of AR-P 6200 (CSAR 62) electron-beam resist. The sample is then installed in a special scanning electron microscope (SEM) that enables CL measurements at cryogenic temperatures. The sample is cooled to 10 K, and the CL signal is mapped over a 20 µm x 20 µm large area with 0.5 µm voxel size to identify the position of bright and isolated QDs that emit at the desired wavelength in the telecom O-band. A corresponding map showing spatially resolved CL in a spectral range of (1300-1305) nm is shown in Fig. 2a). By applying a 2D Gaussian fit, we can select the two circled QDs in Fig. 2a) with a lateral accuracy of about 40 nm. In the following electron beam lithography (EBL) step, deterministic circular mesa structures are written into the resist at low temperature (10 K) in the same SEM system. When the resist is developed, the image areas are cleared and only the EBL-structured mesa structures remain and form an etching mask for the final reactive ion etching in an inductively coupled plasma. For the circular QD micromesas with a gold mirror on the back, we expect a moderately high photon extraction efficiency $\eta$ in a range of 5-10%.[13,15] This parameter can possibly be increased in the future by applying the present manufacturing method to sources based on circular Bragg resonators (CBRs) that promise broadband photon extraction efficiency up to $\eta$ = 95% in the telecom O-band.[16] In fact, our developed device processing is fully compatible with the realization of CBRs and using a soon available commercial in-situ EBL system we plan to implement 1.3 µm QD-CBRs. In this context, it will be interesting to explore the effect of piezo strain tuning on both the QD and cavity properties which could probably be used to induce a piezo-controlled slight ellipticity in the CBGs to realize high-performance QD-SPS with linearly polarized emission under resonant excitation.[17]

Noteworthy, and in contrast to previous work on piezoelectric tuning of QD-micromesas emitting in the 930 nm range,[18] the etching depth used in the present work corresponds to the thickness of the gold-bonded membrane in our design. Thus, the base diameter of the mesa structure is the only cross section to transfer the strain-field effect at the piezoelectric actuator to the QD position. As we discuss below this fact does not have strong impact on the achievable tuning range which is rather limited by non-ideal wafer-bonding.

The properties of the optical device are investigated by means of micro-photoluminescence (µPL), µPL-excitation (µPLE) and photon-correlation spectroscopy at 10 K. The used helium-flow cryostat has an extra-high-voltage feed-through to electrically connect the piezoelectric actuator to a voltage supply and a customized spring holder to facilitate strain-tuning of the QD-micromesas. The deterministic QD mesa structures are optically excited by a continuous-wave (cw) diode laser (785 nm) or a tunable pulsed laser providing ps-pulses at a repetition rate of 80 MHz. The photoluminescence signal is collected with an objective (numerical aperture (NA) = 0.4) and spectrally resolved in a grating spectrometer (spectral resolution ~20 µeV). The photon stream can either be detected by a liquid nitrogen cooled InGaAs-array detector or fiber-based Hanbury Brown and Twiss (HBT) and Hong-Ou-Mandel (HOM) configurations attached to the output slit of the monochromator. The quantum optical configurations include two superconducting nanowire single-photon detectors (SNSPDs) (temporal resolution approx. 50 ps, detection efficiency approx. 80% at 1310 nm).

The device characterization starts by basic µPL measurements under non-resonant cw excitation (785 nm) at 10 K. We identified two photonic structures with bright CX emission lines at 1303.98 nm (QD1) and 1303.78 nm (QD2). The spectral fingerprints of these two QDs, as shown in Fig. 2b), have three characteristic and representative emission lines, which are identified as neutral exciton (X), singly charged exciton (CX) and biexciton (XX) by excitation-power and polarization dependent measurements (see supplementary information (SI) Figs. S1 and S2).

To demonstrate strain-tuning and to quantify the corresponding spectral shift, the QD emission is studied in dependence on the applied voltage to the piezoelectric actuator. The applied voltage was varied from -400 V to +400 V, which corresponds to an electric field $F$ of -13.4 kVcm$^{-1}$ to +13.4 kVcm$^{-1}$, and results in a spectral tuning range of 0.5 nm for the present sample, as shown in Fig. 3. Noteworthy, other samples showed a tuning range up to 4 nm when the device allows for applying higher voltages of up to ±33 kVcm$^{-1}$ (cf. Fig. S3), however, at the risk of sample damage. Comparing these results to values reported in Ref. 18 for a non-completely etched QD-mesa shows that the complete removal of the semiconductor material around the QD-mesa does not significantly reduce the available tuning range since in the electric field range of ± 20 kV/cm the achieved relative spectral shift of (0.24 ± 0.02)% for device B presented in the SI is consistent with (0.20 ± 0.02)% that can be extracted for the device discussed in Ref. 18. In fact, we observe strong device-to-device variations of the available tuning range which indicates non-ideal wafer-bonding using our home-made bonding tool and we expect higher tuning ranges and better reproducibility by using a commercial wafer-bonder in the future.

In addition to the desired wavelength shift of our piezo-controlled QD-micromesas we observe spectral fluctuation over time due to piezoelectric creep, which is most pronounced directly after a voltage change. To illustrate this point Fig. 4a) shows a systematic wavelength change of about 0.05 nm over the first 5 min after setting the piezo voltage. The creep effect decreases significantly after 30 minutes but does not vanish completely over time. This behavior is similar to earlier observations in piezo-controlled QD devices.[8,18,19] In addition, we examined the relative change of the emission energies of X, XX and CX and we found that the binding energy is only slightly influenced, on a scale of 0.3 meV, by the applied strain (see SI Fig. S3). This change of binding energy is significantly smaller than values on the order of meV observed for planar QD samples with obviously better strain transfer.[20] For some QDs the excitonic fine structure splitting (FSS) could be reduced from 40 µeV to 20 µeV (see SI - Fig. S3(d)). Further FSS reduction is not possible in our case. In fact, a second degree of freedom in the piezoelectric actuator would be required to fully symmetrize the distorted confinement potential, being responsible for the FSS.[21]

Long-term spectral stability is one of the most important aspects of a practical single-photon source for quantum communication applications based e.g. on entanglement swapping via Bell-state measurements. Therefore, the effect of creep on the wavelength stability is an important issue to address. One can efficiently stabilize the emission wavelength, as it was reported for a quantum light emitting diode[8] and later also realized for a deterministically fabricated InAs/GaAs QD microlens,[18] by using an active optical feedback (closed-loop) algorithm. In the implemented concept the emission wavelength is monitored by Lorentzian fitting of µPL spectra taken regularly within an integration time of 0.2-1.0 s, depending on the signal intensity. The center position of the fit is compared with the target wavelength and the determined deviation initiates a re-adjustment procedure of the input voltage to shift the emission line back to the set target value. This re-adjustment is based on a PID algorithm.

The implemented wavelength stabilization is demonstrated for the CX emission of QD1 for several PID settings. In this test, we set the proportional gain (PG) to 100, and the integral time (IT) parameter to 0.01 s, 0.05 s and 0.1 s to study its impact on the QD wavelength control after a voltage change applied to the piezo actuator. A graphical representation of the associated results is shown in Fig. 4b), where the following settings are characterized within a time window of 5 minutes for each setting. The corresponding parameter set and achieved wavelength stability are shown in Table 1. Evaluation of the recorded µPL time traces yields that the best spectral stabilization with a standard deviation of the emission wavelength (energy) of 5.6 pm (4.1 µeV) could be achieved for PG = 100 and IT = 0.01 s. Moreover, we found that by increasing PG, the stabilization time decreases. However, in order to maintain the stabilization quality, the IT value must be increased proportionally. We did not observe any clear dependence of the QDs spectral stability on the change of the derivative term of the PID controller.

Next, we demonstrate the CX emission lines of the two different QD-micromesas (QD1 and QD2) to being tuned in resonance and stabilized at this resonance over time by applying the optimized PID settings determined before. The time evolution of the piezoelectrically stabilized CX emission wavelength is presented in Fig. 5a), showing a standard deviation of 12 µeV for QD1 and 6 µeV for QD2, respectively. In Fig. 5b) we present µPL spectra of both QDs tuned into resonance and being stabilized at 1304.05 nm with respect to the CX emission line. The good spectral overlap of the inhomogeneously (by spectral diffusion) broadened emission lines of QD1 and QD2 under active PID control reflects the efficient fine-tuning and spectral control of the QDs via the PID controlled piezo actuator. Ultimately, it has to be demonstrated that high spectral overlap can also be achieved on a scale of the homogenous linewidth which can be probed by HOM experiments on remote sources.

To verify the quantum nature of emission we measured the photon-autocorrelation of the spectrally tunable CX lines of QD1 and QD2. Here, the CX transition is driven at saturation under pulsed non-resonant excitation with 80 MHz repetition rate. In Fig. 6 we present histograms of the correlation events recorded in the HBT configuration representing the second-order autocorrelation function $g^{(2)}(\tau)$ which is used to evaluate the multi-photon suppression at zero time delay. First, the autocorrelation functions of QD1 and QD2 were measured without the influence of the piezo-induced strain-field emitting at different wavelengths, as shown in the upper panels of Fig. 6. The measurements show pronounced antibunching for the central peak at $\tau = 0$ ns, and the evaluation of the experimental data with two-sided mono-exponential fit functions convoluted with the timing response of 50 ps of the SNSPDs used in this experiment, yields $g^{(2)}(0)_{fit} = 0.020^{+0.030}_{-0.020}$ for QD1 and $g^{(2)}(0)_{fit} = 0.052 \pm 0.030$ for QD2. When the electric field is applied to the piezo-actuator and both QDs' CX emissions are set to the resonant wavelength and stabilized by PID for several minutes at 1304.05 nm (the measurements were taken independently), the antibunching behavior is preserved and the corresponding fits yield $g^{(2)}(0)_{fit} = 0.040 \pm 0.030$ for QD1 and $g^{(2)}(0)_{fit} = 0.053 \pm 0.030$ for QD2, confirming that the spectral tuning has no significant impact on the single-photon purity of the QD-micromesas. The achieved results are very promising with regards to the target applications requiring the emission of indistinguishable single photons from remote quantum light sources. However, we would like to point out that the photon indistinguishability of these QD-micromesas is not yet high enough for this purpose. In fact, we examined the TPI of tunable and stabilized CX emission of QD1 under pulsed p-shell excitation in Hong-Ou-Mandel configuration. As we discuss in the SI this experiment yields a HOM visibility of $(16 \pm 8)\%$, a post-selected visibility of $(79 \pm 15)\%$, with a corresponding coherence time of $(470 \pm 85)$ ps and a lifetime of $\tau_1 = (1.54 \pm 0.05)$ ns (see SI Fig. 5 and Fig. 6). These values are in agreement with results achieved recently for non-tunable QD-micromesas[13] which indicates that the optical quality and coherence of the 1.3 µm QD has to be improved in the future, e.g. by optimizing the SRL, in order to fully exploit their potential in advanced quantum communication applications. For instance, the gradual increase of the In-content in the SLR and the overall thickness of the SLR could be varied as an optimization parameter to achieve a smoother transition from the strongly lattice mismatched $In_{0.7}Ga_{0.3}As$ QD layer to the GaAs matrix, thereby reducing the density of (charge trapping) defect states. Alternatively, one could also consider slowing down the GaAs-growth on top of the SLR and try to better „heal" out the roughness introduced by the In-containing layers. HOM-measurements can be a sensitive tool to evaluate the impact of design modifications on the optical and quantum optical properties of the 1.3 µm QDs.

In summary, we demonstrated a deterministically fabricated and tunable QD single-photon source emitting in the telecommunication O-band at 1.3 µm. Our device consisting of a QD micromesa and attached to a piezoelectric actuator allows to fine-tune the QD's emission wavelength by up to 0.5 nm via a voltage-controlled strain-field. The tuning range and an implemented closed-loop PID control system enabled us to stabilize the emission energy with an accuracy (standard deviation) of up to ~4 µeV, and to hold emission lines of two remote QD-micromesas with multiphoton-suppression better than 5% in resonance with a spectral overlap of 60-80%. These results are very promising for the future implementation of long-distance quantum communication networks.

**Supplementary material**

See the supplementary material for information on the identification of excitonic complexes in QD1 and QD2, Hong-Ou-Mandel studies on the photon indistinguishability of QD1, and on optical properties of a second device with larger strain-tuning range.

**Acknowledgements**

This work was funded by the FI-SEQUR project jointly financed by the European Regional Development Fund (EFRE) of the European Union in the framework of the programme to promote research, innovation and technologies (Pro FIT) in Germany within the 2nd Poland-Berlin Photonics Programme, grant No. 2/POLBER-2/2016. Support from the German Research Foundation through CRC 787 "Semiconductor Nanophotonics: Materials, Models, Devices" is also acknowledged. P.M. gratefully acknowledges the financial support from the Polish Ministry of Science and Higher Education within "Mobilnosc Plus – Vedycja" program. We thank T. Heindel for technical support.

**Data availability**

The data that support the findings of this study are available from the corresponding author upon reasonable request.

**References**

[1] H. Wang, J. Qin, X. Ding, M.C. Chen, S. Chen, X. You, Y.M. He, X. Jiang, L. You, Z. Wang, C. Schneider, J.J. Renema, S. Höfling, C.Y. Lu, and J.W. Pan, Phys. Rev. Lett. **123**, 250503 (2019).

[2] P. Kok, C.P. Williams, and J.P. Dowling, Phys. Rev. A - At. Mol. Opt. Phys. **68**, 022301 (2003).

[3] H.J. Kimble, Nature **453**, 1023 (2008).

[4] K. Azuma, K. Tamaki, and H.-K. Lo, Nat. Commun. **6**, 6787 (2015).

[5] J. Yang, C. Nawrath, R. Keil, R. Joos, X. Zhang, B. Höfer, Y. Chen, M. Zopf, M. Jetter, S.L. Portalupi, F. Ding, P. Michler, and O.G. Schmidt, Opt. Express **28**, 19457 (2020).

[6] M. Sartison, L. Engel, S. Kolatschek, F. Olbrich, C. Nawrath, S. Hepp, M. Jetter, P. Michler, and S.L. Portalupi, Appl. Phys. Lett. **113**, 032103 (2018).

[7] Z.H. Xiang, J. Huwer, J. Skiba-Szymanska, R.M. Stevenson, D.J.P. Ellis, I. Farrer, M.B. Ward, D.A. Ritchie, and A.J. Shields, Commun. Phys. **3**, 1 (2020).

[8] R. Trotta, P. Atkinson, J.D. Plumhof, E. Zallo, R.O. Rezaev, S. Kumar, S. Baunack, J.R. Schröter, A. Rastelli, and O.G. Schmidt, Adv. Mater. **24**, 2668 (2012).

[9] B. Höfer, F. Olbrich, J. Kettler, M. Paul, J. Höschele, M. Jetter, S.L. Portalupi, F. Ding, P. Michler, and O.G. Schmidt, AIP Adv. **9**, 085112 (2019).

[10] V.M. Ustinov, N.A. Maleev, A.E. Zhukov, A.R. Kovsh, A.Y. Egorov, A. V. Lunev, B. V. Volovik, I.L. Krestnikov, Y.G. Musikhin, N.A. Bert, P.S. Kop'ev, Z.I. Alferov, N.N. Ledentsov, and D. Bimberg, Appl. Phys. Lett. **74**, 2815 (1999).

[11] M. Paul, F. Olbrich, J. Höschele, S. Schreier, J. Kettler, S.L. Portalupi, M. Jetter, and P. Michler, Appl. Phys. Lett. **111**, 033102 (2017).

[12] S. Strauf, N.G. Stoltz, M.T. Rakher, L. a. Coldren, P.M. Petroff, and D. Bouwmeester, Nat. Photonics **1**, 704 (2007).

[13] N. Srocka, P. Mrowiński, J. Große, M. von Helversen, T. Heindel, S. Rodt, and S. Reitzenstein, Appl. Phys. Lett. **116**, 231104 (2020).

[14] A.R. Clawson, Mater. Sci. Eng. R Reports **31**, 1 (2001).

[15] N. Srocka, A. Musiał, P.-I. Schneider, P. Mrowiński, P. Holewa, S. Burger, D. Quandt, A. Strittmatter, S. Rodt, S. Reitzenstein, and G. Sęk, AIP Adv. **8**, 085205 (2018).

[16] L. Rickert, T. Kupko, S. Rodt, S. Reitzenstein, and T. Heindel, Opt. Express **27**, 36824 (2019).

[17] H. Wang, Y.M. He, T.H. Chung, H. Hu, Y. Yu, S. Chen, X. Ding, M.C. Chen, J. Qin, X.

Yang, R.Z. Liu, Z.C. Duan, J.P. Li, S. Gerhardt, K. Winkler, J. Jurkat, L.J. Wang, N. Gregersen, Y.H. Huo, Q. Dai, S. Yu, S. Höfling, C.Y. Lu, and J.W. Pan, Nat. Photonics **13**, 770 (2019).

[18] M. Schmidt, M. V. Helversen, S. Fischbach, A. Kaganskiy, R. Schmidt, A. Schliwa, T. Heindel, S. Rodt, and S. Reitzenstein, Opt. Mater. Express **10**, 76 (2020).

[19] T. Zander, A. Herklotz, S. Kiravittaya, M. Benyoucef, F. Ding, P. Atkinson, S. Kumar, J.D. Plumhof, K. Dörr, A. Rastelli, and O.G. Schmidt, Opt. Express **17**, 22452 (2009).

[20] F. Ding, R. Singh, J.D. Plumhof, T. Zander, V. Křápek, Y.H. Chen, M. Benyoucef, V. Zwiller, K. Dörr, G. Bester, A. Rastelli, and O.G. Schmidt, Phys. Rev. Lett. **104**, 067405 (2010).

[21] R. Trotta, J. Martín-Sánchez, J.S. Wildmann, G. Piredda, M. Reindl, C. Schimpf, E. Zallo, S. Stroj, J. Edlinger, and A. Rastelli, Nat. Commun. **7**, 10375 (2016).

**Figures**

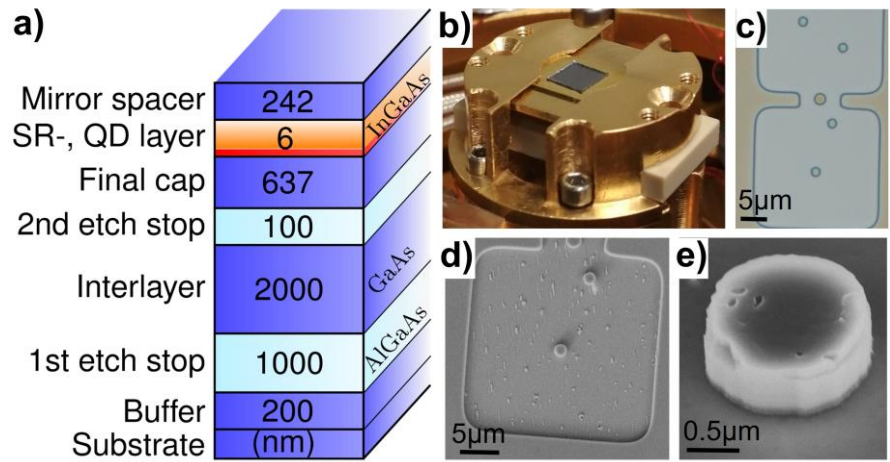

Fig. 1 (a) Sample structure as grown. (b) Sample mounted in the customized spring holder of the cryostat. (c) Optical microscope image of the sample after resist development. (d) SEM image of the processed sample showing the mapping area with the two mesas hosting QD1 and QD2. (e) SEM image of the investigated mesa hosting QD1.

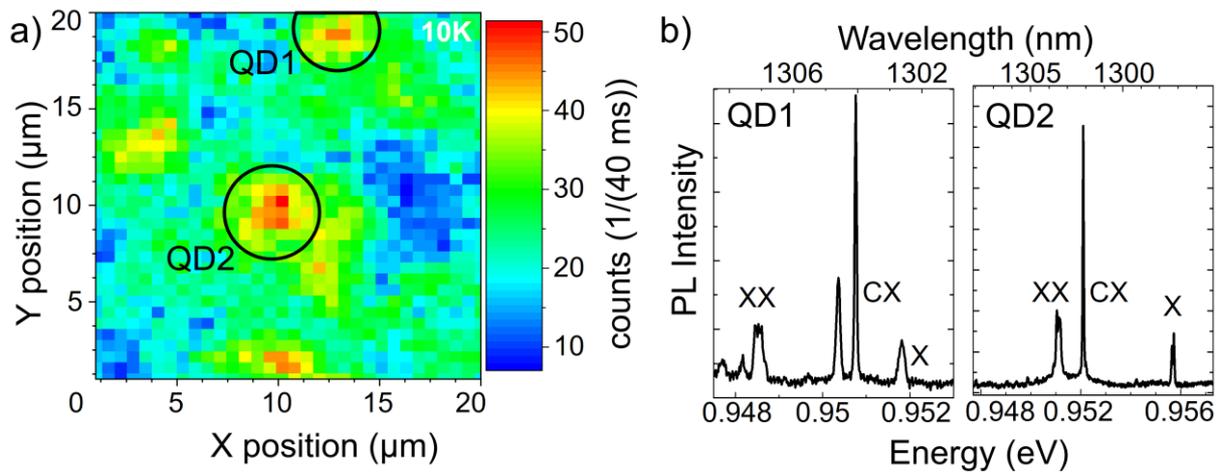

Fig. 2 (a) CL map of the pre-selection step at 10 K with QD1 and QD2 encircled. The spectral range of 1300 nm to 1305 nm is depicted. (b) Corresponding µPL spectra of QD1 and QD2 after mesa processing was completed.

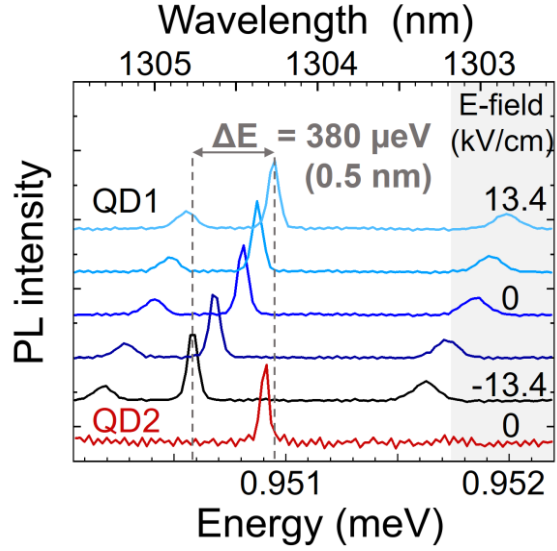

Fig. 3. Energy tuning of the CX emission line of QD1 (black to blue). An energy shift of 380 µeV (0.5 nm) was observed when changing the electric field applied to the piezoelectric actuator from -13.4 kV/cm to +13.4 kV/cm. The CX emission line of QD2 without applied electric field (red) is within the energetic tuning limits of QD1.

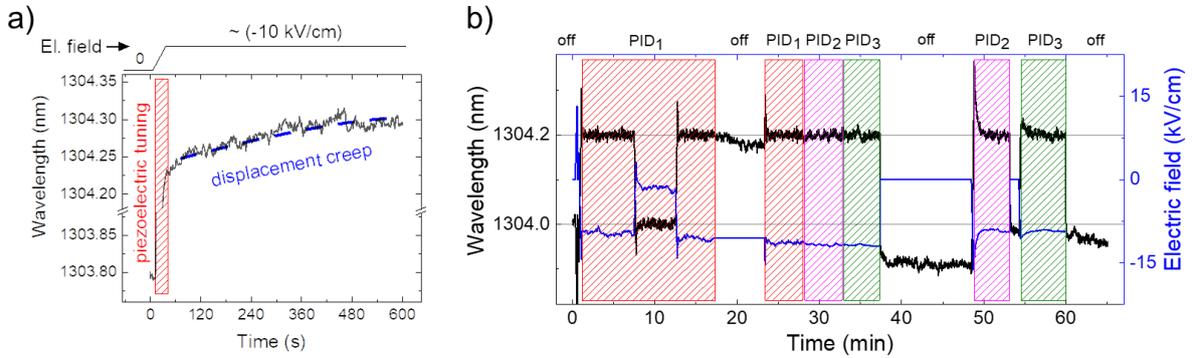

Fig 4. (a) Impact of the piezoelectric actuator's creep on the wavelength of the CX emission line of QD1 over time. (b) PID stabilization tests on QD1 (see Table 1 for the summary).

| PID settings (prop. Gain = 100) | $PID_1$ | $PID_2$ | $PID_3$ |
|---|---|---|---|
| PID integral time (s) | 0.01 | 0.05 | 0.1 |
| standard deviation (µeV) | 4.12 | 4.45 | 5.26 |
| stabilization time (s) | 30 | 50 | 80 |

Table 1. Quantified results of the emission energy stability for various PID integral times evaluated in terms of the standard deviation of emission energy and stabilization time required for the adjustment.

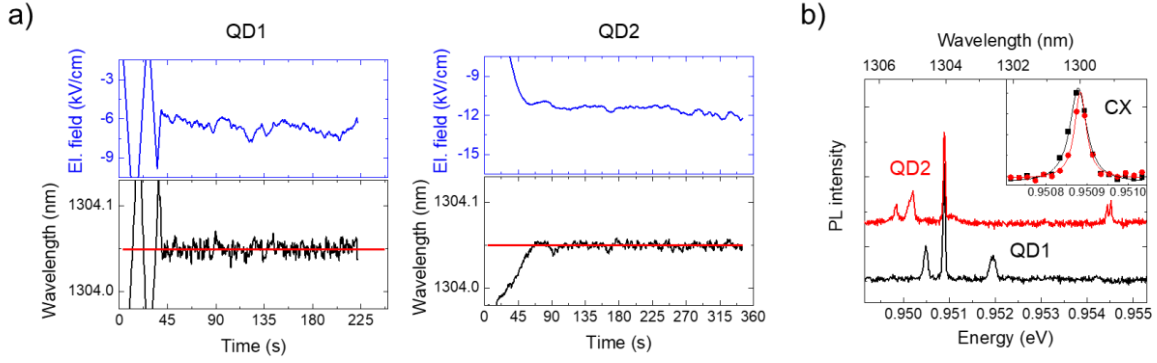

Fig. 5. (b) µPL spectra of QD1 and QD2 when tuned in resonance via applied piezoelectric strain field and (a) emission wavelength traces over time to quantify the stabilization.

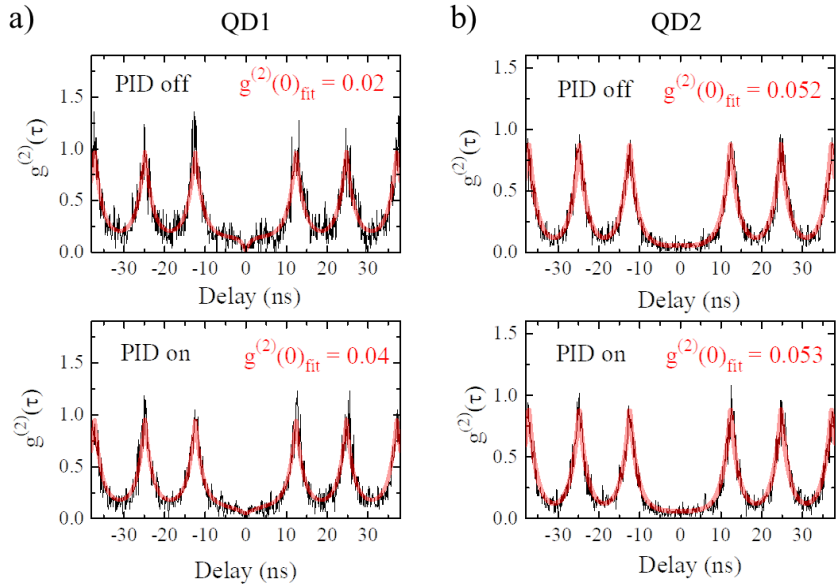

Fig. 6. (a) Autocorrelation experiments performed on the charged excitons of QD1 and (b) QD2 in case of "on" and "off" PID controller which is used for stabilizing the piezo controlled wavelength of the emitters.

# Supplementary Materials: Deterministically fabricated strain-tunable quantum dot single-photon sources emitting in the telecom O-band

N. Srocka[1], P. Mrowiński[1,2], J. Große[1], M. Schmidt[1,3], S. Rodt[1], S. Reitzenstein[1]

[1]*Institut für Festkörperphysik, Technische Universität Berlin, Hardenbergstraße 36, D-10623 Berlin, Germany*

[2]*Laboratory for Optical Spectroscopy of Nanostructures, Department of Experimental Physics, Wrocław University of Technology, Wybrzeże Wyspiańskiego 27, Wrocław, Poland*

[3]*Physikalisch-Technische Bundesanstalt, Abbestraße 2-12, 10587 Berlin, Germany*

## 1. Identification of the excitonic complexes of QD1 and QD2

The excitonic states of the QDs were identified by evaluating power and polarization dependent micro-photoluminescence (µPL) spectra, as shown in Fig. S1 for QD1. The QD was non-resonantly excited by a continuous-wave (cw) diode laser at 785 nm. Already in the low excitation power regime the neutral (X) and charged exciton (CX) are visible. The biexciton (XX) becomes apparent in the high-power regime. With increasing excitation power $P$ the µPL intensity $I$ of X and CX increase proportionally to the excitation power: $I \sim P$. Whereby for the XX the relationship of $P$ and $I$ scales as: $I \sim P^2$. Additionally, to distinguish between the X and CX complex a polarization series in linear basis was performed. The CX complex shows no fine structure splitting (FSS), as expected due to no exchange interactions between electrons and holes. An FSS of approx. 50 µeV was determined for the X state. The line at about 0.95 meV does not show noticeable FSS and most probably corresponds to charged biexcitonic state of QD1. The same analysis was performed for QD2 as presented in Fig. S2. Here the X shows an FSS of 90 µeV.

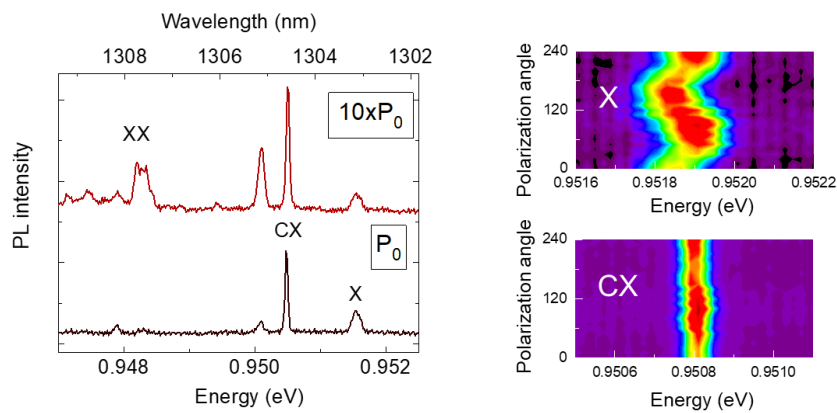

FIG S1. Emission properties of QD 1: µPL spectra in dependence on excitation power (left panel) and polarization angle dependence of X and CX (right panels).

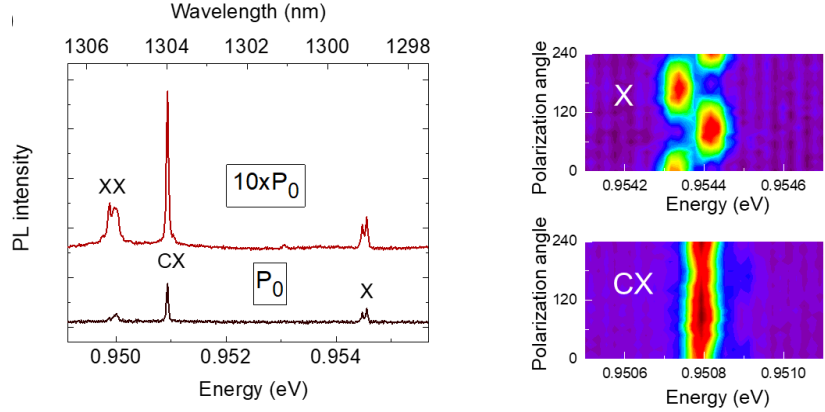

FIG S2. Emission properties of QD 2: µPL spectra in dependence on excitation power (left panel) and polarization angle dependence of X and CX (right panels).

## 2. Optical properties of a second device with larger strain-tuning range

A second device (B), similar to device A (whose results are presented in the main text), was fabricated, using QD material of the same wafer and following the processing steps described in the main text. Fig S3 presents a summary of the main measurements performed on this second device. Analogue to device A, power and polarization dependent µPL series were performed for a selected QD-mesa (QD3) (see Fig. S3 a)-b)) to identify the different excitonic complexes. QD3 shows an FSS of 60 µeV which is similar to the values observed for sample A.

The bonded piezoelectric actuators of device A and B are of the same material (PIN-PMN-PT). Only the piezoelectric actuator as well as the bonded GaAs membrane of device B are twice the size compared to device A. With device B an overall wavelength tuning of 2.9 meV (4.2 nm) could be achieved by tuning the applied electric field in the range of ±33.3 kV/cm (see Fig. S3c)). Comparing only the tuning range for an applied electric field of ±13.3 kV/cm device A allowed for a 0.38 meV (0.5 nm) and device B for a 1.5 meV (2.16 nm) wavelength tuning. Different QDs of one device show similar wavelength shifts for a given applied electric field. However, the maximum tuning range differs for the two devices, which may be due to minor variations in the QD material composition or more likely to processing related divergences. The quality of stress transmission from the piezoelectric actuator to the position of the QDs crucially depends on the gold bonding quality, here the difference in size of the bonded membranes of devices A and B could play a part.

In Fig. S3 d)) we present the FSS and the relative phase of the linearly polarized states of QD3 in dependence of the voltage applied to the piezoelectric actuator. The FSS could be reduced from 60 µeV to a minimum of 20 µeV. Here, the non-zero FSS indicates that the applied strain-field is not accurately aligned to the high symmetry crystal axes [110] or [1-10]. A partial inversion of the polarized states is observed. The polarization of the exciton transition is overall rotated by 65°. Noteworthy, the interdependence between FSS and the relative phase agree well with the theoretical analysis in Ref. [1] under variation of the biaxial strain applied to the QD.

Finally, in Fig. S3 e) the emission energy of X ($E_x$) and the binding energies ($\Delta E_i$) of the CX and XX states are plotted vs. the applied voltage. The binding energies of XX and CX, defined as the difference of emission energies $\Delta E_{XX/CX} = E_X - E_{XX/CX}$, increase under compressive stress (kV/cm < 0) and decrease with tensile stress (kV/cm > 0) in all studied QDs. This observation of an increasing (decreasing) $\Delta E$ can be traced back to the effect of an increase (decrease) of the electron-hole overlap and an associated change of the electron-hole attraction when an in-plane compressive (tensile) biaxial strain is applied.[1]

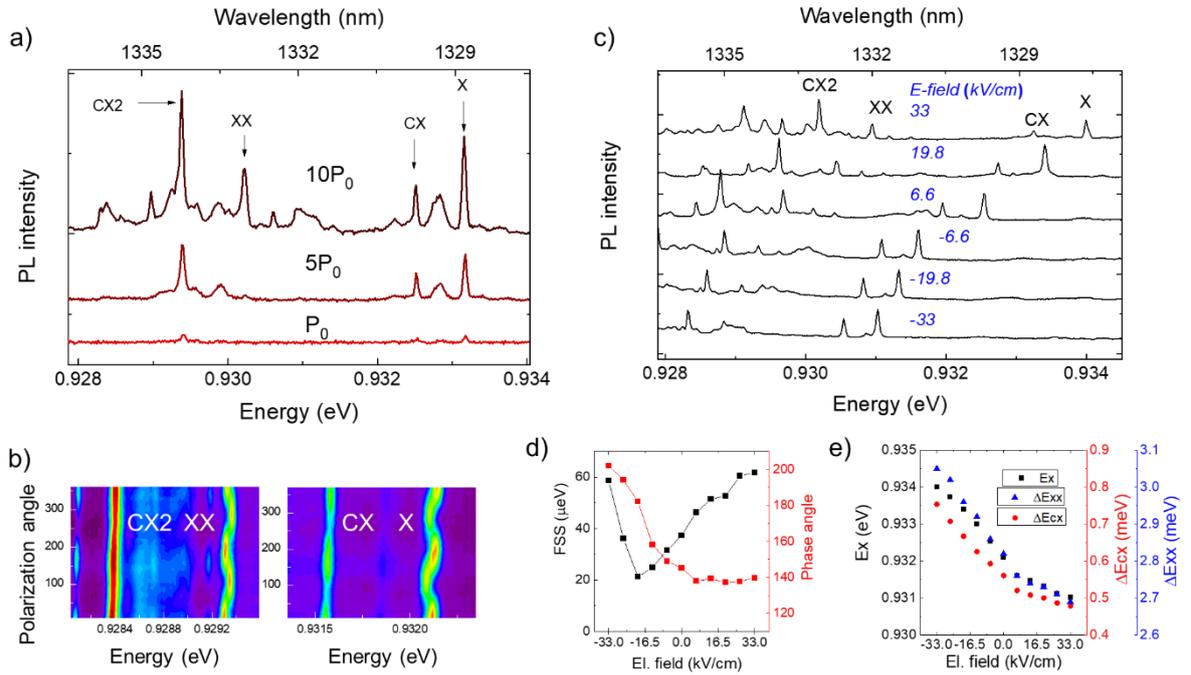

FIG S3. Optical and strain-tuning characteristics of QD3 stemming from sample B. Excitation power (a), polarization (b), and piezo tuning (c) µPL series. The indicated electric field range is associated with an applied electric field of -33 to 33 kV/cm to the piezo-actuator, leading to a tuning range of approx. 4 nm. d) Dependence of the FSS and the polarization phase vs. the applied electric field. e) Exciton energy, and CX and XX binding energies as a function of the applied electric field.

### 3. Hong-Ou-Mandel studies on photon indistinguishability of QD1

To analyze the photon indistinguishability of QD1 we performed pulsed two-photon interference (TPI) measurements using a fiber-based HOM setup with an unbalanced Mach-Zehnder interferometer (MZI) on the excitation side and a complementary MZI on the detection side. The relative optical path delay is 4 ns resulting in clusters of five peaks in the correlation histogram every 12.5 ns for the given laser repetition rate of 80 MHz. The employed fit function is based on a sum of Lorentzians which are temporally shifted from one another in time in steps of about 4 ns. The raw data, the corresponding fit and the fit decomposition are presented in Fig. S4. For a deeper insights into the theoretical resulting histograms see Ref.[2] and for a detailed mathematical description of the applied fit function see Ref. [3].

The measured data is slightly unbalanced and therefore to determine the visibility we use only the central peak area and the areas of the first side peaks of the copolarized HOM histograms. Using the central peak areas $A_0$ and $A_{00}$, where indices "0" and "00" corresponds to the HOM central peak and HOM dip minimum, respectively, the TPI visibility can be determined by[3]:

$$V = \frac{A_0 - (A_0 + A_{00})}{A_0} \times 100\% = (11.7 \pm 6.7)\%,$$

Next, we define the peak area of the first side peak as an average: $A_1 = (A_{1,L} + A_{1,R})/2$ and the TPI visibility is given by[2]:

$$V = \frac{(2/3)A_1 - (A_0 + A_{00})}{(2/3)A_1} \times 100\% = (21 \pm 14)\%.$$

Based on this evaluation we obtain an average value for the photon indistinguishability of $V = (16 \pm 8)\%$. This value agrees with $V = (12 \pm 4)\%$ reported in Ref. 3 for a very similar device with back-side Au mirror but without piezo-element. We attribute the rather low indistinguishability to spectral jitter due to charge fluctuation in the QD environment including the strain-reducing layer. This highlights again that further growth optimization is necessary to improve the photon indistinguishability of the long-wavelength InGaAs QDs, where HOM studies are a very efficient tool to evaluate the quantum-optical properties of these emitters.

To obtain additional insight into the photon indistinguishability and the related coherence properties of emission we apply the above-mentioned fit function, to determine the post-selected visibility. For this purpose we use the value at zero time delay given by the fit function $f(0)$ (see Fig. S4 b)) with respect to the expected maximum in case of distinguishable photons, which could be approximated by the same function neglecting the "00" HOM dip contribution f'(0) (see Fig. S4 b)). Then the post-selected visibility $V_{ps}$ at $\tau = 0$ is given by

$$V_{ps} = \frac{f(0)-f\prime(0)}{f(0)} \times 100\% = (85 \pm 8)\%.$$

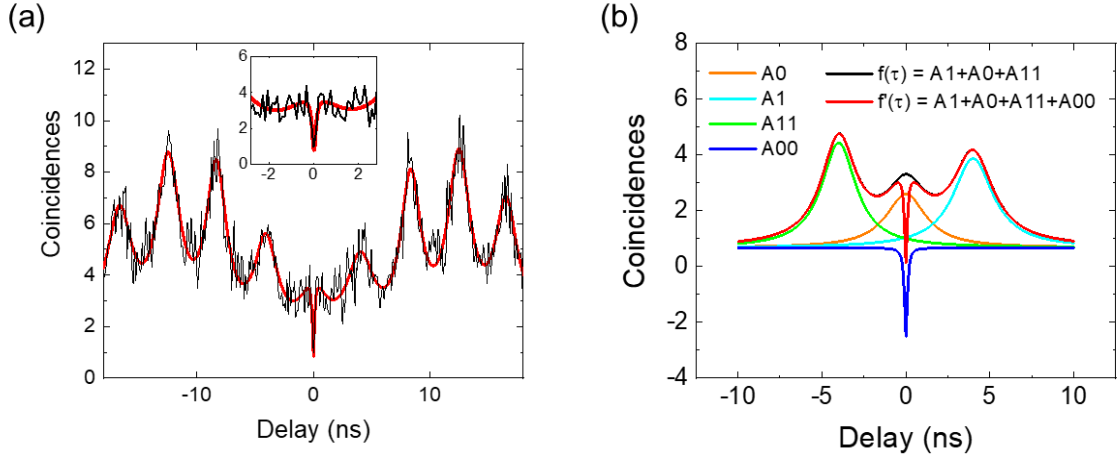

FIG S4. TPI histograms measured for the CX emission of QD1 under pulsed p-shell excitation for co-polarized configuration. The HOM-effect is proven by the highly reduced coincidences at zero-time delay. The red trace corresponds to the applied fit function based on a sum of Lorentzian peak functions. In part (b) the decomposition into individual peaks is shown in order to determine post selected visibility from f(τ=0) and f'(τ=0). In the fit we consider an uncorrelated background of 0.67.

A further analysis of the central dip of the HOM histogram for parallel polarized photons allows for an estimation of the coherence time $\tau_c$, which is characterised by the width of this central dip (cf. Fig S5). The central dip is fitted by a double exponential decay function and convoluted with the response function of the detectors (time jitter ~ 50 ps). The first side peaks at $\pm 4$ ns were as well considered for the fit, in the form of their previous decomposed Lorentzian functions. For the mathematical description see Ref. [2]. With the visibility $V$ fixed to 1 and the three free fitting parameters lifetime of spontaneous emission $\tau_1$, the dephasing time $\tau_{deph}$ and the background $y_0$ we obtained $\tau_1 = (1.54 \pm 0.05)$ ns and $\tau_{deph} = (0.55 \pm 0.08)$ ns for a background $y_0 \ll 0.01$. The correspondent fit is depicted in Fig. S5. Taking dephasing into account and therefore following the relation $\frac{1}{\tau_{deph}} = \frac{1}{\tau_c} - \frac{1}{2\tau_1}$, we obtain a coherence time of $\tau_c = (0.47 \pm 0.08)$ ns. Noteworthy, the extracted lifetime is consistent with a value $\tau_1 = (1.58 \pm 0.05)$ ns obtained independently by time-resolved µPL measurements.

The post-selection approach naturally leads to higher TPI visibility which is essentially limited by the temporal resolution of the used single-photon detectors. In order to enlarge the low visibility without post-selection and to produce high-purity triggered indistinguishable single photons, it is necessary to reduce the dephasing rate and timing jitter e.g. by s-shell resonant excitation or by reducing charge noise by electric control in contacted devices. Furthermore, additional optimization of the strain-reducing layer to reduce the number of charged defect states, leading to spectral jitter, could improve the quantum optical quality of the devices.

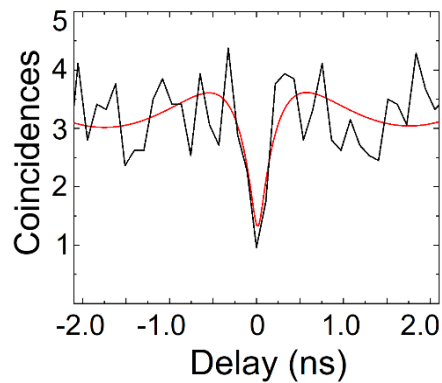

FIG S5. Close-up of the HOM histogram for parallel polarized photons. The TPI dip is fitted by a double exponential decay function taking the response function of the SNSPDs into account, see text for details.

**References**

[1] F. Ding, R. Singh, J.D. Plumhof, T. Zander, V. Křápek, Y.H. Chen, M. Benyoucef, V. Zwiller, K. Dörr, G. Bester, A. Rastelli, and O.G. Schmidt, Phys. Rev. Lett. **104**, 067405 (2010).

[2] A. Thoma, P. Schnauber, M. Gschrey, M. Seifried, J. Wolters, J.-H.H. Schulze, A. Strittmatter, S. Rodt, A. Carmele, A. Knorr, T. Heindel, and S. Reitzenstein, Phys. Rev. Lett. **116**, 033601 (2016).

[3] N. Srocka, P. Mrowiński, J. Große, M. von Helversen, T. Heindel, S. Rodt, and S. Reitzenstein, Appl. Phys. Lett. **116**, 231104 (2020).